\documentclass[prb,twocolumn,showpacs,preprintnumbers]{revtex4}
\usepackage{graphicx}
\begin{document}

\title{Order to disorder transition in the XY-like quantum magnet ${\bf Cs_2CoCl_4}$ induced by noncommuting applied fields}

\author{M. Kenzelmann,$^{1}$ R. Coldea,$^{1,2,3}$ D.~A. Tennant,$^{1,3}$ D. Visser,$^{4,5,6}$ M.
Hofmann,$^{7}$ P. Smeibidl,$^{7}$ and Z. Tylczynski$^{8}$}

\affiliation{(1) Oxford Physics, Clarendon Laboratory, Oxford OX1
3PU, United Kingdom\\ (2) Oak Ridge National Laboratory, Oak
Ridge, Tennessee 37831\\ (3) ISIS Facility, Rutherford Appleton
Laboratory, Oxon OX11 0QX, United Kingdom\\(4) NWO-EW, ISIS
Facilty, Rutherford Appleton Laboratory, Chilton, Didcot, OX11
0QX,UK\\(5)Department of Physics, University of Warwick, Gibbet
Hill Road,Coventry CV4 7AL, UK\\(6) IRI, Technical University
Delft,Mekelweg 15,  2629JB Delft, The Netherlands\\(7)
Hahn-Meitner-Institut, BENSC, 14109 Berlin, Germany\\(8) Institute
of Physics, Adam Mickiewicz University, Umultowska
85, 61-614 Poznan, Poland}%


\begin{abstract}
We explore the effects of noncommuting applied fields on the
ground-state ordering of the quasi-one-dimensional spin-1/2
XY-like antiferromagnet Cs$_2$CoCl$_4$ using single-crystal
neutron diffraction. In zero field interchain couplings cause
long-range order below $T_N$=217(5) mK with chains ordered
antiferromagnetically along their length and moments confined to
the ($b,c$) plane. Magnetic fields applied at an angle to the XY
planes are found to initially stabilize the order by promoting a
spin-flop phase with an increased perpendicular antiferromagnetic
moment. In higher fields the antiferromagnetic order becomes
unstable and a transition occurs to a phase with no long-range
order in the ($b,c$) plane, proposed to be a spin liquid phase
that arises when the quantum fluctuations induced by the
noncommuting field become strong enough to overcome ordering
tendencies. Magnetization measurements confirm that saturation
occurs at much higher fields and that the proposed spin-liquid
state exists in the region 2.10$<H_{SL}<$2.52 T $\parallel$ $a$.
The observed phase diagram is discussed in terms of known results
on XY-like chains in coexisting longitudinal and transverse
fields.
\end{abstract}

\date{\today}
\maketitle

\section{Introduction}

A promising area for the study of zero-temperature quantum phase
transitions is that of quantum spin systems in noncommuting
applied magnetic fields \cite{sachdev99}. Such noncommuting terms
introduce quantum fluctuations into the $T=0$ ground state which,
for large enough fields, can completely disorder the system. Such
a situation has been studied in considerable detail in the
three-dimensional (3D) Ising ferromagnet LiHoF$_{4}$ in a
transverse magnetic field \cite{bitko96}. However, due to its high
dimensionality, the system behaves in a mean-field like way. In
this paper we consider a one-dimensional (1D) quantum magnet and
the effects of a noncommuting field on its ground state.\par

The scenario that we investigate experimentally is that of the 1D
spin-1/2 XXZ model (${\cal H}_{XXZ}(\Delta )$) in a noncommuting
field (${\cal H}_{APP}$), given by the Hamiltonian,
\begin{equation}
{\cal H} ={\cal H}_{XXZ}(\Delta )+{\cal H}_{APP},
\end{equation}
where,
\begin{equation}
{\cal H}_{XXZ}(\Delta ) = J\;\sum_{i}
\left(S_{i}^{x}S_{i+1}^{x}+S_{i}^{y}S_{i+1}^{y} + \Delta
S_{i}^{z}S_{i+1}^{z}\right)\, \label{Eq_Hamil_CsCoCl}
\end{equation}
and
\begin{equation}
{\cal H}_{APP} =\sum_{i}B^{x}S_{i}^{x}+B^{z}S_{i}^{z},
\end{equation}
and where $J>0$ is the antiferromagnetic (AF) exchange constant,
and $\Delta <1$\ is the anisotropy parameter (a $g_{x,z}\mu_B$
factor is incorporated in the magnetic field $B^{x,z}$). The
operators $S^{\alpha }$, where $\alpha =x,y,z$,\ are the usual
spin operators for spin-1/2, and the applied field term does not
commute with the exchange Hamiltonian $[{\cal H}_{XXZ}(\Delta
),{\cal H}_{APP}]\neq 0$ for $\left| \Delta \right| \neq 1$ and
$B^{x}\neq 0$. Indeed, the system we shall study,
Cs$_{2}$CoCl$_{4}$, is an excellent realization of ${\cal
H}_{XXZ}(\Delta )$ where $\Delta =0.25$, and therefore should be
well approximated by the famous XY model ($\Delta =0$)\
\cite{LSM}.\par

The physics of the ${\cal H}_{XXZ}(\Delta )$ model at $\Delta =0$
is well known in the absence of a magnetic field i.e.
$B^{x}=B^{z}=0$. It is that of a 1D noninteracting Fermi gas as
shown by the Jordan-Wigner transformation \cite{Jordan_Wigner}.
The effect of a field along $z$, $B^{z}\neq 0$, (commuting) is
trivial; it acts as a chemical potential and changes the filling
level of the fermions in the chain inducing magnetization along
$z$. Below a critical field, $B_{C}^{z}$, representing the field
of complete saturation of the chain, the excitation spectrum
remains gapless and the $T=0$ correlation functions fall
algebraically as
power laws. In contrast, the physics for the case of a field along $x$, $%
B^{x}\neq 0,$ $B^{z}=0$, is neither trivial nor widely known.\par

\begin{figure}
\begin{center}
  \includegraphics[height=4.5cm,bbllx=126,bblly=334,bburx=471,
  bbury=518,angle=0,clip=]{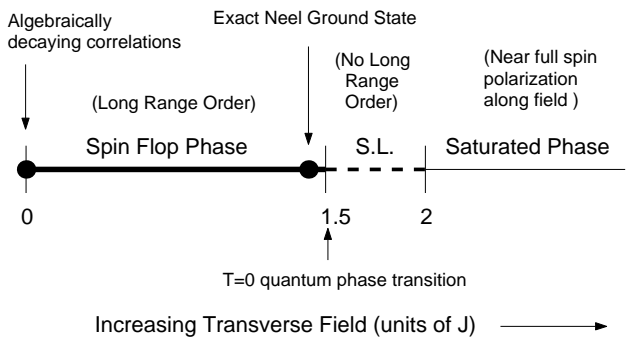}
  \caption{Schematic representation of the phases of the ground
    state of the XY-model as a function of applied transverse field
    as proposed by Kurmann {\it et al.}\protect\cite{Kurmann81,Kurmann82}
    (see text for details). In the absence of a field the
    correlation functions decay as a
    power law. Small magnetic fields induce perpendicular long-range
    ordered antiferromagnetism and the ordering is characterized as
    a spin flop phase. At a field $\sqrt{2}J$ the classical N\'eel
    state is the exact $T=0$ ground state. The antiferromagnetism is
    rapidly suppressed at higher fields by the quantum fluctuations
    induced by the noncommuting field. Above about $1.5J$ long range
    order is destroyed by these fluctuations and the ground state is
    characterized as a spin liquid (S.L.) state with exponentially
    decaying correlations in the spin components perpendicular to the
    field. Above a crossover field of about 2$J$ nearly all the spin
    moments are aligned along the field direction and the physics here
    is characterized as a saturated phase.}
  \label{Fig_theoretical_phase_diagram}
\end{center}
\end{figure}

The action of noncommuting field $B^{x}$ on ${\cal H}_{XXZ}(0)$
has been considered theoretically by Kurmann {\it et al.}
\cite{Kurmann81,Kurmann82}. From these studies it was found that
the in-plane field has two effects:\ 1)\ it breaks the U(1)
symmetry of the XY-model to a lower, Ising-like, symmetry which
causes the ground state to long-range order (LRO) at $T=0$ into a
spin-flop type N\'eel state. In fact at a special coupling,
$B^{x}=\sqrt{2}J$ the spin-flop N\'eel state is the exact $T=0$
ground state. 2)\ The second effect is to introduce quantum
fluctuations into the system. At high fields this causes a phase
transition to occur where the fluctuations become large enough to
destroy the LRO altogether. This disordering field is below that
where the system reaches its saturation point. This phase
transition is therefore a nontrivial quantum phase transition
through a quantum critical point with the noncommuting field as a
control parameter. Fig.~\ref{Fig_theoretical_phase_diagram} shows
a schematic outline of the physics of the XY-model in a transverse
field.\par

The theoretical studies \cite{Kurmann81,Kurmann82} suggest that
this behavior is generic to a wide class of magnets in
noncommuting fields, and a disordered spin-liquid (S.L.) phase is
expected for all $H_{XXZ}(| \Delta | \neq 1)$ in equation (1).
Spin liquid phases such as this are generally gapped (evidenced by
the exponentially decaying correlations in the zero-temperature
ground state) and therefore robust against small perturbations. We
therefore expect the effects of nonzero $\Delta$ and interchain
coupling to modify the transition fields in
Fig.~\ref{Fig_theoretical_phase_diagram} but not to change its
qualitative content.

In this paper we present a detailed study of the ground state
ordering of the quasi-1D XY-like antiferromagnet ${\rm
Cs_2CoCl_4}$ in noncommuting fields from zero to well above
saturation. The crystalline and magnetic properties of ${\rm
Cs_2CoCl_4}$ and the experimental tool of neutron diffraction will
be introduced in Section~\ref{Sec_experiment}.\par

Section~\ref{Sec_results} presents the observed commensurate
magnetic structure and its field dependence. The structure is that
of magnetic chains which are ordered antiferromagnetically along
their lengths with ordered moments lying in the ($b$,$c$) plane.
The experimental results show that an applied field initially
further stabilizes the ordered structure (spin-flop phase), but as
the field increases fluctuations induced by field noncommutation
cause a sharp transition to a new phase which we propose to be the
spin-liquid phase discussed above. Study of the ferromagnetic
component (\ref{Subsec_ferromagnetic_moment}) confirms that the
saturation field is higher still, and that the proposed spin
liquid phase exists in the region $2.10 < H_{SL} < 2.52$ Tesla
$\parallel$$a$.

An extended discussion of our results is given in
Section~\ref{Sec_Disc}. The observation of commensurate order
($\bbox k$=(0,1/2,1/2)) in ${\rm Cs_2CoCl_4}$ is related to the
effective Ising-type interchain coupling
(\ref{Subsec_commensurate}) and compared with the quasi-elastic
results of Yoshizawa {\it et al.}\cite{Yoshizawa} 
A microscopic study of the ordering is presented in
\ref{Subsec_mean_field} based on a mean-field analysis of the
ground-state energy. The magnitude of the ordered moment is
established in Section~\ref{Subsec_magnitude}. Finally, in
Section~\ref{Subsec_field_dependence} the magnetization curve is
critically examined and compared with known results for
anisotropic spin chains in noncommuting applied fields.

\section{Experimental Details}
\label{Sec_experiment}

\subsection{Properties of ${\bf Cs_{2}CoCl_{4}}$}
\bigskip
${\rm Cs_{2}CoCl_{4}}$ has been proposed as a spin-1/2 quasi-1D
XY-like antiferromagnet with chains running along the
$b$-direction. Heat-capacity measurements\cite{Algra} showed a
broad maximum (characteristic of low-dimensional systems) around
$T\simeq 0.9\;{\rm K}$ and the overall temperature dependence
agreed well with numerical predictions\cite{Katsura} for a $S$=1/2
XY-like AF chain with exchange coupling $J=0.23(1)\;\mathrm{meV}$.
A small lambda-type anomaly observed in the specific-heat at
$T_{N}=222\;\mathrm{mK}$ was interpreted\cite{Algra} as indicating
a phase transition to a magnetically ordered phase caused by small
couplings between chains. The in-plane magnetic susceptibility
\cite{McElearney,Duxbury} also indicated magnetic ordering below
$222\;\mathrm{mK}$ and the temperature dependence between
$40\;\mathrm{mK}$ and $4.2\;\mathrm{K}$ was in excellent agreement
with numerical calculations for AF XXZ chains with $\Delta=0.25$
in Eq~\ref{Eq_Hamil_CsCoCl}.\par

Quasi-elastic neutron scattering experiments\cite{Yoshizawa}
showed that between $T=0.3$ and $0.6\;\mathrm{K}$ the critical
scattering of ${\rm Cs_{2}CoCl_{4}}$ is sheet-like perpendicular
to the $b$ axis, thus confirming the proposed quasi-1D character
of the magnetic properties. Interestingly, the critical scattering
is driven partly to incommensurate positions due to competing
inter-chain interactions. Those earlier diffraction experiments
\cite{Yoshizawa} concentrated on the critical scattering at
temperatures above the ordering transition of $0.22\;\mathrm{K}$
inferred by macroscopic measurements. As the ordering in this
material promises to have some unusual and challenging features,
we undertook detailed neutron diffraction experiments to determine
the magnetic structure and the behavior in applied noncommuting
magnetic fields at temperatures much below the proposed ordering
transition of $0.22\;\mathrm{K}$.\par

\begin{figure}
\begin{center}
  \includegraphics[height=7cm,bbllx=100,bblly=150,bburx=550,
  bbury=670,angle=0,clip=]{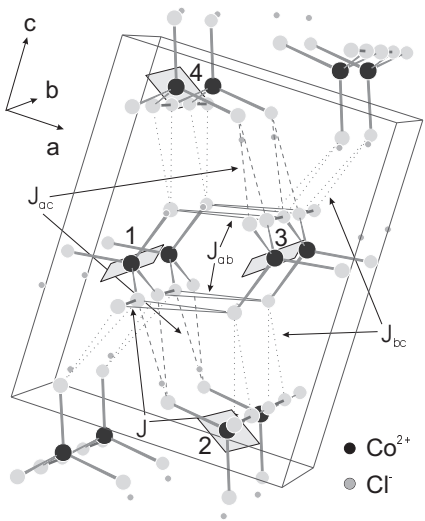}
  \caption{Crystal structure of ${\rm Cs_2CoCl_4}$. The figure shows
  $12$ Co$^{2+}$ ions, each surrounded by a distorted tetrahedron
  of Cl$^{-}$-ions. The Co$^{2+}$ ions interact mainly via an
  AF super-exchange interaction $J$ along the $b$
  axis, forming AF spin chains which interact weakly
  via super-exchange interactions $J_{ab}$, $J_{ac}$ and $J_{bc}$
  as explained in the text. The shaded rectangular planes indicate
  possible orientations of the XY easy planes
  following Ref.~\protect\onlinecite{Figgis}.}
  \label{crystal_structure}
\end{center}
\end{figure}

${\rm Cs_2CoCl_4}$ crystallizes in the orthorhombic (and
nonsymmorphic) space group \textit{Pnma} ($D^{16}_{2h}$, No. 62)
\cite{Pnam}. The crystal structure is shown schematically in
Fig.~\ref{crystal_structure}. The lattice parameters at 0.3 K
are\cite{Figgis_Reynolds,Yoshizawa} $a=9.71\,$\AA \, ,
$b=7.27\,$\AA \, and $c=12.73\,$\AA. The magnetic ions, Co$^{2+}$
with spin $\tilde{S}=\frac{3}{2}$, occupy site 4$c$ in the unit
cell at positions\cite{Figgis_Reynolds,Figgis_Reynolds_comment}
\begin{eqnarray}
&(1):\hspace{0.5cm} (0.235, 0.25, 0.422) & \nonumber \\& (2):
\hspace{0.5cm} (0.735, 0.25, 0.078) & \nonumber \\& (3):
\hspace{0.5cm} (0.765, 0.75, 0.578) & \nonumber \\& (4):
\hspace{0.5cm} (0.265, 0.75, 0.922) &\, . \nonumber
\end{eqnarray}Neighboring spins interact via a
superexchange interaction involving a bridge of two Cl$^{-}$-ions
with the path Co$^{2+}$-Cl$^{-}$-Cl$^{-}$-Co$^{2+}$. The shortest
Cl$^{-}$-Cl$^{-}$ distance is between neighbors along the $b$-axis
separated by $3.63\,$\AA. As this is close to twice the ionic
radius of Cl a sizeable overlap of electron wave-functions
contributing to the exchange integral is expected.\par

Among other possible exchange paths are $J_{ac}$ between sites 1
and 2 ($d_{{\rm Cl}^{-}-{\rm Cl}^{-}}$= 4.05 \AA), $J_{ab}$
between sites 1 and 3 ($d_{{\rm Cl}^{-}-{\rm Cl}^{-}}$= 4.04 \AA),
and $J_{bc}$ between sites 1 and 4 ($d_{{\rm Cl}^{-}-{\rm
Cl}^{-}}$=4.01 \AA). Since the overlap of the electronic
wave-functions decreases very rapidly (approximately
exponentially) with distance, these exchange paths are expected to
yield much smaller interactions than the coupling $J$ along $b$
such that ${\rm Cs_2CoCl_4}$ can be regarded as a system of
weakly-coupled spin chains along the $b$ axis.\par

Each Co$^{2+}$ ion is tetrahedrally coordinated by Cl$^{-}$
ligands. Small distortions from a perfect tetrahedron lead to a
splitting of the $\tilde{S}$=$\frac{3}{2}$ orbital singlet state
into two Kramers doublets with a separation $2D$=1.3(1) meV. The
magnetic exchange energy is much lower than the inter-doublet
separation and therefore only the lowest-lying doublet states
participate in the low-energy dynamics at low temperatures ($T \ll
D$). Projecting the Heisenberg exchange between the true spins
onto the lowest-lying doublet of $|\pm\frac{1}{2}\rangle$ states
gives an effective spin-1/2 Hamiltonian with XY-like exchange,
${\cal H}_{XXZ}$ in Eq.~\ref{Eq_Hamil_CsCoCl} with $\Delta$=0.25.

Rotations of the CoCl$_4$ tetrahedra in the unit cell lead to
different orientations of the XY easy plane between sites (1,3)
and (2,4) and give the $b$-axis as the only common in-plane
direction for all sites. The main distortion of the CoCl$_4$ unit
from a perfect tetrahedron is due to one of the four Cl$^{-}$ ions
being rotated by several degrees around the $b$-axis with respect
to the central Co$^{2+}$ ion and previous studies\cite{Figgis}
proposed that the normal to the XY easy plane ($z$-axis) bisects
this large angle. This gives the $z$-axis direction
($\sin\beta$,0,$\cos\beta$) on sites (1,3) and
($-\sin\beta$,0,$\cos\beta$) on sites (2,4) with
$\beta=-38.8^{\circ}$. The shaded rectangular planes in
Fig.~\ref{crystal_structure} show the XY planes in this case.
Another possibility is that the $z$ axis is along the vector
connecting the central Co$^{2+}$ ion with the Cl$^{-}$ ion rotated
most and in that case $\beta$=+19.4$^{\circ}$.\par

\subsection{Experimental method}
\label{Subsec_experimental_method} A high-quality $6.45\;{\rm g}$
single crystal of ${\rm Cs_{2}CoCl_{4}}$ was grown from solution.
The crystal was aligned with its
$(0,k,l)$ plane in the horizontal scattering plane and was cooled
to temperatures between $T=80$ and $250\;{\rm mK}$ using an Oxford
Instruments dilution refrigerator insert placed inside a vertical
7 Tesla superconducting magnet.

Neutron diffraction measurements were made at the Hahn-Meitner
Institut in Berlin, Germany. The two-axis crystal diffractometer
E6 was employed with an incident energy of $E_i$=14.72 meV and
with a pyrolytic graphite (PG) monochromator in double-focusing
mode to increase the neutron flux at the sample position. The
scattered neutrons were counted in a 20$^{\circ}$ wide BF$_3$
detector bank with position sensitivity along the horizontal
direction giving 200 channels. This gave an angular resolution of
0.1$^{\circ}$ in the total scattering angle $2\Theta$. The
intensities of nuclear and magnetic Bragg reflections were
measured as a function of both $2\Theta$ and the sample rotation
angle $\Psi$, thus constructing full two-dimensional (2D) maps of
the scattering intensity in the ($\Psi$, $2\Theta$) plane. This
allowed simultaneous coverage of both magnetic signal and
background. Typical counting times were 5 minutes per 2D map to
determine the total integrated intensity, and 40 seconds for a
$2\Theta$ scan at the peak center to extract the peak intensity. A
two-dimensional Gaussian with adjustable rotation of the main axes
of the ellipsoid gave a good account of the observed peak line
shapes in both $\Psi$ and $2\Theta$ and the intensities of the
Bragg reflections were obtained from least-square fits to the
experimental data to reduce the sum of discrepancies $\chi^2$.\par

\section{Experimental results}
\label{Sec_results}
\subsection{Magnetic order in zero field}
\label{Subsec_mag_order} 
\begin{figure}
\begin{center}
\includegraphics[height=5.3cm,bbllx=75,bblly=265,bburx=482,
    bbury=570,angle=0,clip=]{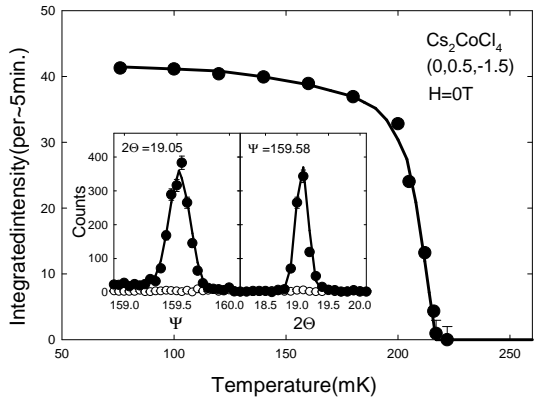}
    \caption{Integrated intensity of the antiferromagnetic
    $(0,0.5,-1.5)$ reflection vs. temperature in zero external field.
    Intensity units are the same as in Table~\ref{Table_Braggpeaks}.
    The solid line is a guide to the eye. The inset shows the Bragg
    peak intensity as a function of $\Psi$ (left) and $2\Theta$ (right)
    at temperatures below (solid circles) and above (open circles) the
    transition temperature $T_N$=217 mK. Solid lines come from a
    two-dimensional fit to the data in the ($\Psi$,$2\Theta$) plane as
    described in the text.} \label{E6_temperature}
\end{center}
\end{figure}

Upon cooling below $T_N$=217 mK extra
Bragg reflections were observed at the commensurate
$(0,n+0.5,m+0.5)$ reciprocal lattice positions with $n$ and $m$
integers, indicating a transition to a magnetically long-range
ordered state. Fig.~\ref{E6_temperature} shows the temperature
dependence of the $(0,0.5,-1.5)$ AF reflection showing the onset
of order below $T_N$. The extracted transition temperature
$T_N$=217 mK is consistent with that inferred from specific-heat
and susceptibility measurements.\cite{Algra,Duxbury}\par

The observed magnetic reflections are associated with a magnetic
ordering wavevector $\bbox{k}=(0,1/2,1/2)$. To determine the
magnetic structure we first used group theory to identify the spin
configurations consistent with the wavevector $\bbox{k}$ for the
given crystal symmetry, and second we compared the structure
factor of possible spin configurations with the experimentally
observed magnetic Bragg peak intensities.\par

The group theory analysis and determination of the
symmetry-allowed basis vectors is presented in Appendix A. After
comparison of the data with the possible eigenvectors, we find
that the observed structure belongs to the $\Gamma^{10}$
irreducible representation with eigenvector $\phi^{10}$ given in
Eq.~\ref{rep10}. This eigenvector has 6 degrees of freedom
corresponding to the three components of the moments $\bbox{m}_1$
and $\bbox{m}_3$. Using spherical coordinates these can be written
as
$\bbox{m}_1=M_1(\sin{\vartheta_1},\cos{\vartheta_1}\cos{\phi_1},
\cos{\vartheta_1}\sin{\phi_1})$ and
$\bbox{m}_3=M_3(\sin{\vartheta_3},\cos{\vartheta_3}\cos{\phi_3},
\cos{\vartheta_3}\sin{\phi_3})$. For $\vartheta_{1,3}=0$ spins are
in the ($b,c$) plane and $\phi$ is the azimuthal angle with the
$b$-axis.\par

\begin{figure}
\begin{center}
    \includegraphics[height=10.5cm,bbllx=98,bblly=122,bburx=538,
    bbury=804,angle=0,clip=]{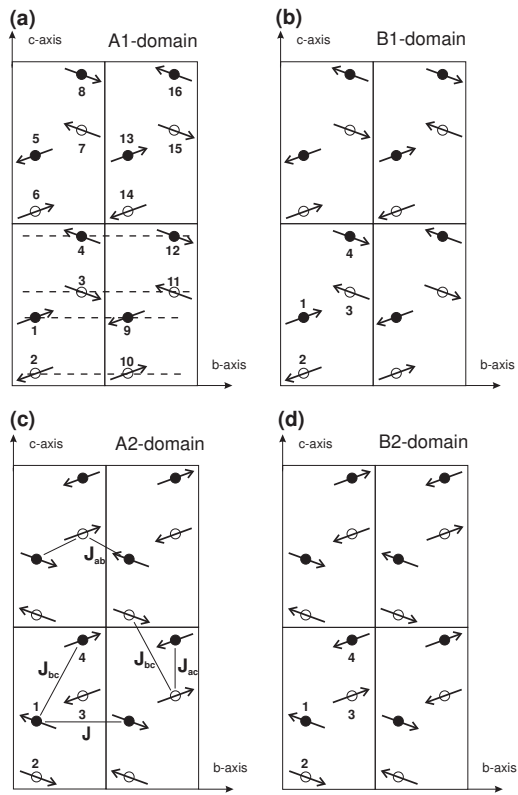}
    \caption{Magnetic structure of Cs$_2$CoCl$_4$. Spins
    (indicated by arrows) order antiferromagnetically along
    chains (shown by dashed lines in (a)). Ordered moments
    are contained in the ($b,c$) plane and make a small
    angle with the $b$-axis (see text for details).
    Relative ordering of the chains leads to degenerate
    domains (a)-(d) belonging to the same irreducible
    representation $\Gamma^{10}$ with  eigenvectors
    given in Eq.~\protect\ref{rep10}. Labels 1-16 in
    (a) indicate the 16 spins in the magnetic unit cell
    (1-4 label the four atoms in the chemical unit cell
    shown in Fig.~\protect\ref{crystal_structure}). Solid
    and open circles are Co$^{2+}$-ions with height along
    the $a$-axis close to 0.25 and 0.75, respectively.}
    \label{Fig-magn-structure}
\end{center}
\end{figure}

Fig.~\ref{Fig-magn-structure}(a) shows a pictorial representation
of the $\phi^{10}$ eigenvector in Eq.~\ref{rep10} in the special
case of ordered spins contained in the ($b,c$) plane, making a
small angle with the $b$-axis ($\phi_1=-\phi_3$) and having equal
magnitude on all sites ($|\bbox{m_1}|=|\bbox{m_3}|$). The
structure can be described in terms of antiferromagnetic chains
along $b$ with a certain ordering pattern between adjacent chains.
Starting with the basic structure shown in
Fig.~\ref{Fig-magn-structure}(a) other distinct domains shown in
Fig.~\ref{Fig-magn-structure}(b)-(d) can be constructed by
changing the sign of either the $b$- or the $c$- spin components,
or the relative phase between chains 1 and 3 (for details see
Appendix A).\par

We find that the measured Bragg intensities can be consistently
described by an equal population of A and B domains of
$\Gamma^{10}$. The best fit of the model to the magnetic
intensities (for details see Appendix) is shown in
Fig.~\ref{Fig_Braggpeaks} and in Table~\ref{Table_Braggpeaks}, and
gives a good description of the experimental data. A single-domain
structure of either A- or B-type as defined in Appendix A and
shown in Fig.~\ref{Fig-magn-structure} is unable to account for
the results.\par

\begin{figure}
\begin{center}
\includegraphics[height=5.2cm,bbllx=76,bblly=272,bburx=486,
    bbury=570,angle=0,clip=]{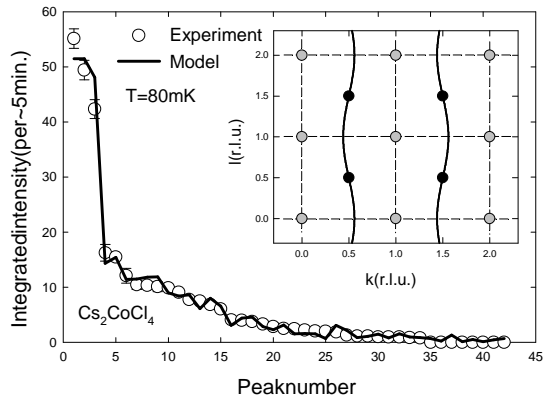}
    \caption{Experimentally observed integrated magnetic Bragg peak
    intensities (open circles) fitted to the model for the magnetic
    structure described in the text (solid line). Intensity units are
    the same as in Fig~\protect\ref{E6_temperature}. The horizontal axis
    indicates the number of the Bragg peaks in the list given in
    Table~\ref{Table_Braggpeaks}. Inset: Schematic diagram
    of the $(b,c)$ reciprocal space indicating the positions of magnetic
    Bragg peaks (solid circles) observed in the present experiment
    in the ordered phase below $T_N$=217(5) mK. Grey circles indicate
    reciprocal lattice positions in the ($b,c$) plane and undulating
    solid lines mark the positions of the sheets of quasi-elastic
    magnetic scattering observed in earlier experiments\protect\cite{Yoshizawa}
    at higher temperatures ($T$$>$0.3 K) above the transition to the
    ordered phase.}\label{Fig_Braggpeaks}
\end{center}
\end{figure}

Assuming spin moments confined to the ($b,c$) plane
($\vartheta_{1}=\vartheta_{3}=0$) and equal ordered moments on all
sites ($|\bbox{m}_1|=|\bbox{m}_3|$) the best fit results are
$\phi_{1}=15(5)^{\circ}$, $\phi_{3}=-15(5)^{\circ}$ and fraction
of A-domain $\alpha=0.48(3)$ with the sum of discrepancies
$\chi^2=2.55$. The obtained value for $\alpha\sim 1/2$
demonstrates that domains A and B occupy the sample in equal
parts. It will be shown in Section~\ref{Subsec_mean_field} that
those two domains have the same mean-field exchange energy and are
thus expected to occur with the same probability. Taking
$\bbox{m}_1$ out of the ($b,c$) plane by choosing a series of
non-zero values for the out-of-plane angle $\vartheta_{1}$ and
fitting $\vartheta_{3}$ leads to worse agreement with the observed
intensities, so within the accuracy of the experiment it was
concluded that magnetic moments are mostly contained in the
($b,c$) plane.\par

\begin{table}
\caption{Experimentally observed magnetic Bragg peak intensities
compared to the calculated intensities based on the model
explained in the text. The magnetic Bragg reflections are in the
order of decreasing observed intensity. A visual comparison
between experiment and model is made in Fig.~\ref{Fig_Braggpeaks}}
\begin{ruledtabular}
\begin{tabular}{ccccccc}
&Number&{$\bbox{Q}$}=($h$,$k$,$l$) & Exp. Int. & Calc. Int. &\\
\hline &1&(0,0.5,1.5) & 55.1(1.8) &  51.4 &\\ &2&(0,-0.5,-1.5) &
49.4(1.8) &  51.4 &\\ &3&(0,0.5,-1.5) & 42.4(1.7) &  44.8 &\\
&4&(0,-0.5,-4.5) & 16.2(1.5) &  15.1 &\\ &5&(0,0.5,-4.5) & 15.5(1)
&  15.5 &\\ &6&(0,0.5,0.5) & 12.1(1.4) & 9.7 &\\ &7&(0,1.5,1.5) &
10.4(0.8) & 10.1 &\\ &8&(0,0.5,-0.5) & 10.3(0.6) & 8.1 &\\
&9&(0,1.5,-1.5) & 10.1(0.9) & 8.4 &\\ &10&(0,-1.5,-4.5) & 9.9(0.2)
& 10.1 &\\ &11&(0,1.5,-4.5) & 9.1(0.2) & 8.8 &\\ &12&(0,-0.5,-3.5)
& 7.8(0.2) & 8.8 &\\ &13&(0,-0.5,-2.5) & 7.5(0.4) & 6.0 &\\
&14&(0,0.5,-3.5) & 6.9(0.2) & 8.0 &\\ &15&(0,0.5,-2.5) & 6.1(0.2)
& 6.0 &\\ &16&(0,0.5,-6.5) & 4.1(0.4) & 3.5 &\\ &17&(0,2.5,-4.5) &
4.0(0.2) & 4.3 &\\ &18&(0,1.5,-3.5) & 3.8(0.4) & 4.1 &\\
&19&(0,-0.5,-6.5) & 3.3(0.4) & 3.4 &\\ &20&(0,1.5,2.5) & 2.8(0.6)
& 2.4 &\\ &21&(0,2.5,-1.5) & 2.5(0.4) & 2.2 &\\ &22&(0,3.5,-4.5) &
2.4(0.4) & 1.8 &\\ &23&(0,2.5,3.5) & 2.2(0.6) & 1.9 &\\
&24&(0,1.5,-0.5) & 2.0(0.6) & 0.84 &\\ &25&(0,3.5,-3.5) & 2.0(0.6)
& 0.61 &\\ &26&(0,2.5,1.5) & 1.9(0.4) & 3.0 &\\ &27&(0,1.5,-2.5) &
1.3(0.4) & 2.0 &\\ &28&(0,2.5,-2.5) & 1.14(0.2) & 0.64 &\\
&29&(0,-0.5,-5.5) & 1.14(0.4) & 1.2 &\\ &30&(0,1.5,0.5) & 1.0(0.2)
&  1.1 &\\ &31&(0,2.5,2.5) & 0.95(0.4) & 0.8 &\\ &32&(0,2.5,-3.5)
& 0.94(0.2) & 1.5 &\\ &33&(0,0.5,-5.5) & 0.84(0.3) & 1.1 &\\
&34&(0,2.5,-2.5) & 0.79(0.6) & 0.64 &\\ &35&(0,2.5,0.5) & 0(0.4) &
0.40 &\\ &36&(0,3.5,0.5) & 0(0.4) & 0.18 &\\ &37&(0,3.5,-1.5) &
0(0.4) & 0.83 &\\ &38&(0,4.5,-0.5) & 0(0.4) & 0.08 &\\
&39&(0,4.5,-1.5) & 0(0.4) & 0.39 &\\ &40&(0,4.5,0.5) & 0(0.4) &
0.09 &\\ &41&(0,4.5,1.5) & 0(0.4) & 0.5 &\\ &42&(0,-1.5,-5.5) &
0(0.4) & 0.8 &\\
\\\end{tabular}\end{ruledtabular}\label{Table_Braggpeaks}
\end{table}

\begin{figure}
\begin{center}
\includegraphics[height=5.1cm,bbllx=30,bblly=265,bburx=582,
    bbury=570,angle=0,clip=]{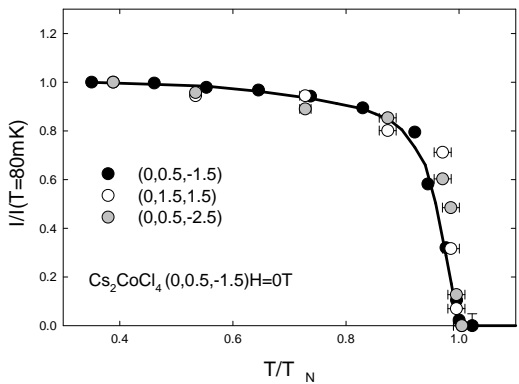}
    \caption{Zero-field reduced integrated intensity $I/I_{T=80\mathrm{mK}}$
    for three AF Bragg reflections as a function
    of reduced temperature $T/T_N$. The solid line is a guide
    to the eye.}
    \label{Fig_temp3AF_peaks}
\end{center}
\end{figure}

Fig.~\ref{Fig_temp3AF_peaks} compares the temperature dependence
of the reduced intensity $I/I_{T=80\mathrm{mK}}$ of three AF
reflections. Here $I_{T=80\mathrm{mK}}$ is the observed intensity
at base temperature $T=80\;\mathrm{mK}$. The three reflections
have essentially the same temperature dependence, indicating that
the magnetic structure is unchanged between 80 mK and $T_N$.\par

\subsection{Magnetic order in applied field}
\label{Subsec_mag_order_field} 
\begin{figure}
\begin{center}
  \includegraphics[height=5.3cm,bbllx=50,bblly=265,bburx=512,
  bbury=570,angle=0,clip=]{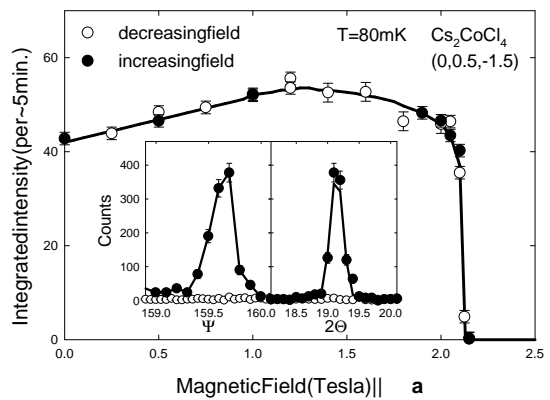}
  \caption{Integrated intensity
  of the $(0,0.5,-1.5)$ AF reflection vs. applied field at $T=80\;\mathrm{mK}$.
  Intensity units are the same as in Table~\ref{Table_Braggpeaks}.
  The solid line is a guide to the eye. The inset shows the Bragg
  peak intensity as a function of $\Psi$ (left) and $2\Theta$ (right)
  in applied fields below (solid circles) and above $H_c$ (open
  circles). The solid lines correspond to a fit to the data in the
  ($\Psi$, $2\Theta$) plane as described in the text.}
  \label{Fig_field_1AF_80mK}
\end{center}
\end{figure}

The effect of an applied field on
the magnetic order was investigated for fields along $a$. A
spin-flop phase arises (spins cant out of the ($b,c$) plane
towards the field axis) evidenced by a perpendicular
antiferromagnetic order coexisting with a ferromagnetic moment
along the field direction. Fig.~\ref{Fig_field_1AF_80mK} shows the
field dependence of the $(0,0.5,-1.5)$ reflection measuring the
antiferromagnetic component. The intensity increases with
increasing field, reaches a maximum around $H \sim
1.4\;\mathrm{T}$ and then it abruptly drops to zero at
$H_c=2.10(4)\;\mathrm{T}$ in a sharp, near-first order phase
transition. Throughout the spin-flop phase ($0<H<H_c$) the
ordering wave-vector was constant at the commensurate position
$(0,1/2,1/2$).\par

The transition at $H_c$ was measured both with increasing and
decreasing field, and no measurable hysteresis effect was observed
as shown in Fig.~\ref{Fig_field_1AF_80mK}. At $3\;\mathrm{T}$, no
AF or incommensurate reflections were observed along symmetry
direction in the ($b,c$) plane with an intensity larger than
$1.5\%$ of the zero-field intensity of the strong $(0,0.5,-1.5)$
reflection, equivalent to half the average background level. The
absence of magnetic Bragg peaks in the ($b,c$) plane suggests that
the phase immediately above the critical field $H_{c}$ is a
disordered phase (no LRO) stabilized by the applied magnetic fields
{\it c.f.} the disordered SL phase in Fig. 1.\par

\begin{figure}
\begin{center}
  \includegraphics[height=5.3cm,bbllx=50,bblly=265,bburx=512,
  bbury=570,angle=0,clip=]{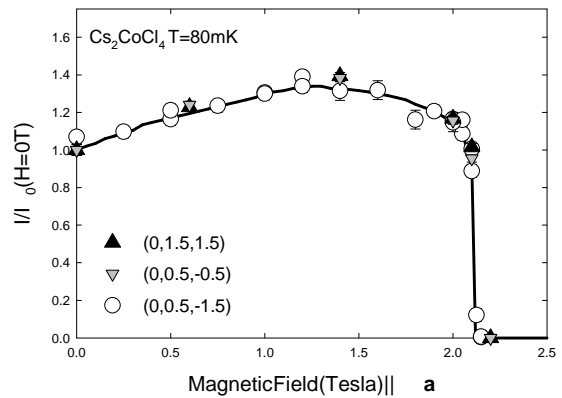}
  \caption{Reduced integrated magnetic Bragg peak intensity, $I/I_0$,
  as a function of field for three AF reflections at $T=80\;\mathrm{K}$.
  The data for $(0,1.5,1.5)$ and $(0,0.5,-0.5)$ were measured with
  increasing field, while most of the data for $(0,0.5,-1.5)$ was
  measured with decreasing field as shown in
  Fig.~\protect\ref{Fig_field_1AF_80mK}. The solid line is a guide to
  the eye.}\label{Fig_3AFpeaks}
\end{center}
\end{figure}

Fig.~\ref{Fig_3AFpeaks} compares the field dependence of the
reduced intensity $I/I_0$ of three AF reflections at
$T=80\;\mathrm{mK}$, where $I_0$ is the zero-field integrated
intensity. The field dependence coincides showing that the
magnetic structure formed by the antiferromagnetic moments in the
$(b,c)$ plane is unchanged throughout the spin flop phase up to
the critical field $H_c$. Furthermore, it shows that the initial
increase in the AF Bragg peak intensities is not due to a
rearrangement of the moments in a different spin configuration but
arises from an increase in the magnitude of the
antiferromagnetically ordered moment. This effect is attributed to
the suppression of zero-point quantum fluctuations by the applied
field which allows more of the available spin moment to order.\par

\begin{figure}
\begin{center}
  \includegraphics[height=5.3cm,bbllx=50,bblly=265,bburx=512,
  bbury=570,angle=0,clip=]{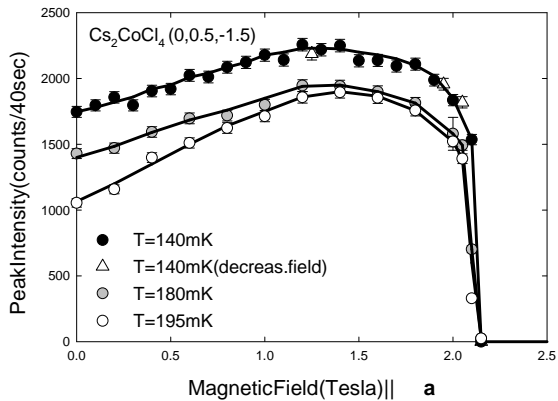}
  \caption{Peak intensity of the $(0,0.5,-1.5)$
  AF reflection vs. applied field at three different
  temperatures. Intensity units are the same as in
  Fig.~\protect\ref{Fig_temp_in_field}. The solid
  lines are guides to the eye.}
  \label{Fig_field-180-195}
\end{center}
\end{figure}

The effect of the applied field initially suppressing fluctuations
and stabilizing the antiferromagnetic order is even more
pronounced at elevated temperatures where the zero-field moment
value is further reduced by thermal fluctuations.
Fig.~\ref{Fig_field-180-195} shows that the antiferromagnetic
moment increases significantly in applied field to reach a maximum
around 1.4 T and only at much higher fields it collapses in a
sharp transition at $H_c$=2.10(4) T (nearly $T$-independent up to
195 mK=0.9 $T_N$).

\begin{figure}
\begin{center}
  \includegraphics[height=5.3cm,bbllx=50,bblly=265,bburx=512,
  bbury=570,angle=0,clip=]{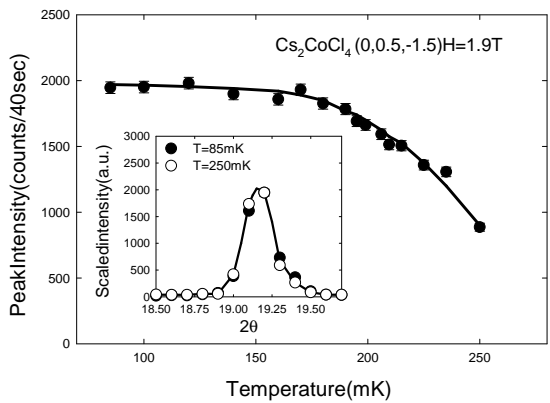}
  \caption{Peak intensity of the $(0,0.5,-1.5)$ AF reflection
  vs. temperature in fixed applied field $H=1.9\;\mathrm{T}$.
  Intensity units are the same as in
  Fig.~\ref{Fig_field-180-195}. The solid line is a guide to the
  eye. Inset: observed intensity of the $(0,0.5,-1.5)$ reflection
  as a function of scattering angle $2\Theta$ at two
  temperatures $T=80$ and $250\;\mathrm{mK}$. The
  $250\;\mathrm{mK}$ data was scaled to match the peak intensity
  of the $80\;\mathrm{mK}$ data for direct comparison. The solid
  line is a Gaussian fit.}
  \label{Fig_temp_in_field}
\end{center}
\end{figure}

Consistent with the above observation, scans in temperature at
intermediate fields (below $H_c$) observe that the transition
temperature is higher than in zero field, as expected for a
structure with an increased ordered moment (stabilized by the
field) that has a larger mean-field energy and is thus more stable
against thermal fluctuations. This is shown in
Fig.~\ref{Fig_temp_in_field} by measurements of (0,0.5,-1.5) peak
intensity in a field of $H$=1.9 T. The intensity decreases slower
with increasing temperature than in zero field [see
Fig.~\ref{E6_temperature}] and the order is stable beyond the
zero-field transition temperature $T_N$=217(5) mK and disappears
only at a much higher temperatures above the range covered by this
experiment, extrapolated to $T_N(H$=1.9 T$)$ $\sim$ 300(20) mK.
The long-range coherence of the structure is unchanged in the
whole measured temperature range as evidenced by the same angular
width of $2\Theta$ scans at the two extreme temperatures shown in
the inset of Fig.~\ref{Fig_temp_in_field}.\par
\begin{figure}
\begin{center}
  \includegraphics[height=5.3cm,bbllx=69,bblly=268,bburx=480,
  bbury=568,angle=0,clip=]{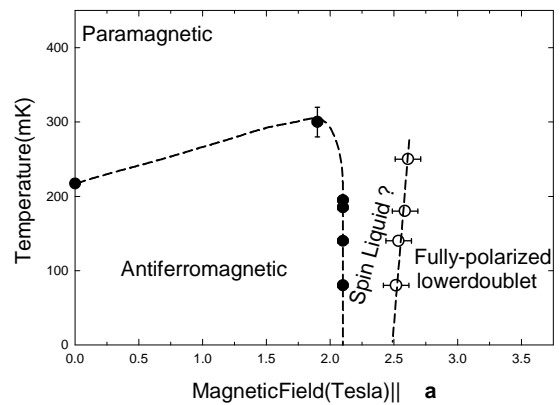}
  \caption{Schematic magnetic phase diagram of ${\rm Cs_2CoCl_4}$ as a
  function of magnetic field and temperature. Solid symbols denote
  transitions where the AF LRO in the ($b,c$)
  plane disappears, see Figs.~\ref{E6_temperature}(\ref{Fig_temp_in_field})
  and \ref{Fig_field_1AF_80mK}(\ref{Fig_3AFpeaks}).
  Open symbols mark cross-over fields in the magnetization curve
  (see Fig.~\ref{Fig_3F-peaks}) identified with near-saturation
  of the lower doublet magnetization. Dashed lines
  are guide to the eye.}
  \label{Fig_phase-diagram}
\end{center}
\end{figure}

\subsection{Temperature-field phase diagram}

In Fig.~\ref{Fig_phase-diagram} we show an ($H \parallel a$,$T$)
phase diagram of ${\rm Cs_2CoCl_4}$ based on the measurements
described above. At zero field, ${\rm Cs_2CoCl_4}$ orders below
$T_N$=217(5) mK with a commensurate wave-vector
$\bbox{k}=(0,1/2,1/2)$. The chains are ordered
antiferromagnetically along their length and the moments are
contained in the ($b,c$) plane at a small angle with the $b$-axis.
Magnetic fields applied along the $a$-axis initially stabilize the
antiferromagnetic order by suppressing fluctuations. This is
directly observed both in the increase of the perpendicular
antiferromagnetic moment in applied field (see
Figs.~\ref{Fig_3AFpeaks}-\ref{Fig_field-180-195}) and also in the
increase in the transition temperature at finite field (compare
Figs.~\ref{E6_temperature} and \ref{Fig_temp_in_field}). At higher
fields the order becomes unstable and above $H_c$=2.10(4)
T$\parallel$ $a$ it collapses in a sharp, near-first order
transition to a phase with no long-range magnetic order in the
($b,c$) plane, possibly a spin-liquid phase. Based on 1) the
absence of order in the ($b,c$) plane, 2) nonsaturation of the
magnetization, see Section~\ref{Subsec_ferromagnetic_moment}, and
3) the expected order-disorder transition driven by the large
fluctuations arising from field noncommutation, we identify the
phase above the critical field $H_c$ with the spin-liquid (S.L.)
state predicted to occur below saturation in
Fig.~\ref{Fig_theoretical_phase_diagram}.\par

\subsection{Ferromagnetic moment as a function of field}

\label{Subsec_ferromagnetic_moment} 
\begin{figure}
\begin{center}
  \includegraphics[height=5.5cm,bbllx=77,bblly=265,bburx=485,
  bbury=569,angle=0,clip=]{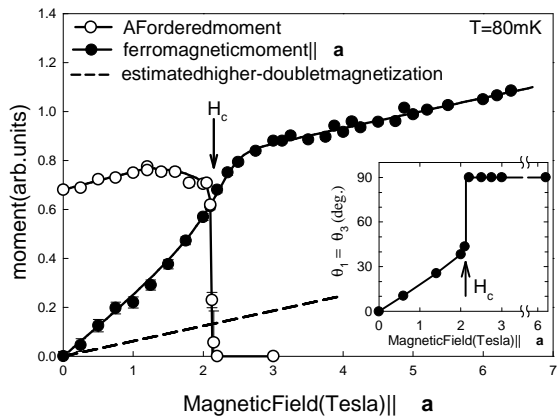}
  \caption{Ordered moments as a function of applied field
  along $a$ at $T$=80 mK. Open circles show the perpendicular
  ordered moment and solid symbols indicate the ferromagnetic
  ($\parallel$$a$) component.
  The solid lines are guides to the eye and the dashed line shows
  the estimated partial contribution to the magnetization due to
  polarizing the higher-doublet states alone, assuming a constant
  susceptibility vs. field. The inset shows the canting angle
  $\theta$ made by the total ordered moment with the ($b,c$) plane
  (above $H_c$=2.1 T it is assumed that the spin components
  perpendicular to the field axis are not ordered and therefore
  $\theta$=90$^{\circ}$).}
  \label{Fig_moments}
\end{center}
\end{figure}

\begin{figure}
\begin{center}
  \includegraphics[width=8cm,bbllx=75,bblly=263,bburx=500,
  bbury=680,angle=0,clip=]{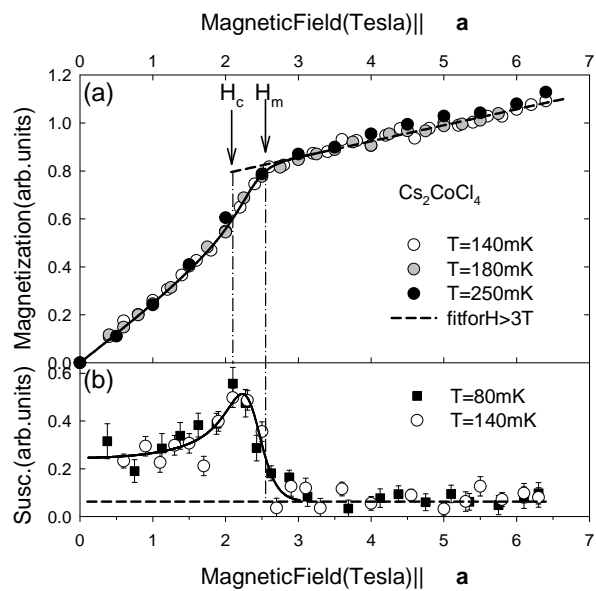}
  \caption{(a) Magnetization vs. field along the $a$-axis
  at various temperatures $T$=140, 180 and 250 mK. The
  solid line is a guide to the eye and
  the sloping dashed line shows a fit to a linear behavior at fields
  above $\sim$3 T. (b) Susceptibility vs. field at $T$=80
  and 140 mK. The solid line is guide to the eye (the horizontal
  dashed line shows the approximated partial contribution from
  the higher-doublet states). $H_c$ is the critical field where
  the antiferromagnetic order in the ($b,c$) plane
  disappears and $H_m$ is the cross-over field above which the
  magnetization has approached the high-field near-linear behavior.}
  \label{Fig_3F-peaks}
\end{center}
\end{figure}

The ferromagnetically-ordered
moment was determined from the intensity of the
(011) reflection. The resulting magnetization curve
at $T$=80 mK is shown in Fig.~\ref{Fig_moments}. The ferromagnetic
moment increases over the whole range of the measurements up to
6.4 T and two regimes can be identified: a low- and a high-field
region separated by a cross-over around $H_m$=2.52$\pm$0.06 T
above which the rate of increase of the magnetization is
significantly reduced. This cross-over is best illustrated in a
plot of the differential susceptibility ($\chi=\partial M
/\partial H$) in Fig.~\ref{Fig_3F-peaks}(b) which shows
significantly reduced values above $H_m$ (of order 4$\pm$1
compared to zero field). This cross-over behavior is typical of
spin-3/2 systems with two energetically-separated Kramers doublets
(for details see Section~\ref{Subsec_field_dependence}): in
applied field the lower-lying doublet is saturated first at a
field $H_m$ above which the magnetization curve shows a large
decrease in susceptibility as only higher-doublet states can still
be polarized.

The cross-over field $H_m$ (defined experimentally as the field
where the magnetization is within less than 5$\%$ of the linear
high-field behavior) is plotted in Fig.~\ref{Fig_phase-diagram}
(open symbols). Note that the antiferromagnetic order in the
($b,c$) plane is suppressed at $H_c$=2.10(4) T much below the
lower-doublet saturation field identified with $H_m$=2.52$\pm$0.06
T ($T$=80 mK). The absence of long-range order for $H_c<H<H_m$ is
not due to thermal fluctuations as this field range remains finite
after extrapolation to $T$=0 as shown in
Fig.~\ref{Fig_phase-diagram}. This region is thus a quantum
disordered phase induced by the applied field and is thus
consistent with the proposed spin-liquid phase in the schematic
phase diagram in Fig.~\ref{Fig_theoretical_phase_diagram}.\par

\section{Discussion}\label{Sec_Disc}
In the previous section we have established the magnetic structure
of Cs$_2$CoCl$_4$ and the effects of a noncommuting field on its
ground state. Here we discuss relevant microscopic mechanisms that
may give rise to the observed ordered structure and find that a
simple (nearest neighbor) mean-field picture can not explain all
observed features. We then estimate the ordered moment and find a
significant reduction from the available moment, indicating the
presence of strong quantum fluctuations. Finally, we relate the
observed magnetic phase diagram to the minimal model of the
XY-chain in a noncommuting field.\par

\subsection{Commensurate vs. incommensurate order}
\label{Subsec_commensurate} The observed ordering wave-vector
condenses out of the diffuse scattering measured by Yoshizawa {\it
et al.}\cite{Yoshizawa} in the disordered phase at 0.3 K just
above $T_N$ (inset of Fig.~\ref{Fig_Braggpeaks}), consistent with
the expectation that order should arise at wavevectors where
paramagnetic fluctuations are strong. The measured diffuse
scattering\cite{Yoshizawa} showed systematic incommensurate
modulations along the chain direction attributed to competing
interchain interactions. In isotropic systems, a mean-field
picture for such frustrated couplings would predict
incommensurate, spiral spin order along the chains
\cite{Yoshizawa}. The observed order however occurs at the
commensurate, antiferromagnetic wavevector $\bbox{k}$=(0,1/2,1/2)
where interchain frustration effects cancel out and chains behave
as decoupled.\par

In Cs$_2$CoCl$_4$ the tendency to form spiral order is suppressed
because XY planes of neighboring chains are not parallel but make
a large relative angle 2$\beta$ resulting in an Ising-like
frustrated interchain coupling between the XY-like 1D chains.
Considering only chains 1 and 4 in the chemical unit cell with XY
spins the energy of a helical order in the easy planes of the two
chains at a pitch to minimize frustration for the $b$-axis spin
components is $E_{helix}=S^2[-J-|\cos2\beta|J_{bc}^2/(2J)]$. This
spiral structure becomes degenerate with the simple AF N\'{e}el
order (spins along $b$) in the limit of orthogonal easy planes
$2\beta$=90$^{\circ}$ (close to the actual situation in
Cs$_2$CoCl$_4$). In the proximity of this limit of small effective
frustration other effects may stabilize the observed commensurate
N\'{e}el-type order and possibilities include: (1) zero-point
quantum fluctuations could introduce non-linear terms in the
free-energy expansion which may lift the classical degeneracy and
promote ordering at the decoupling point $\bbox{k}$=(0,1/2,1/2),
(2) further neighbor couplings such as between sites 1 and 5, or 1
and 6, although believed to be small could potentially stabilize
the antiferromagnetic order, and (3) other effective Ising-type
anisotropies arising from spin-orbit coupling or crystal field
effects may favor spin ordering close to the $b$-axis with the
lowest energy achieved for an antiferromagnetic-type arrangement
(constant moment on each site) as opposed to other incommensurate
structures. We have also considered dipolar couplings and found an
increased energy by 10$^{-5}$ meV per spin for the observed order
compared to a ferromagnetic-type arrangement along the $c$-axis
and thus it was concluded that dipolar effects could not explain
the observed structure.\par

\subsection{Mean-field analysis}
\label{Subsec_mean_field} To identify the exchange couplings
involved in stabilizing the observed magnetic structure we
calculate its energy in the mean-field approximation. We find
simple energetic arguments to explain: 1) why the observed
magnetic structure belongs to the $\Gamma^{10}$ irreducible
representation, 2) why the ordered moments are confined to the
($b,c$) plane and 3) why a multi-domain structure occurs.\par

We first assume an isotropic Heisenberg exchange such that the
three spin components along $x$, $y$ and $z$ ($a$-, $b$- and
$c$-directions) could be treated separately. The observed ordering
wave-vector $\bbox{k}$=(0,1/2,1/2) indicates a doubling of the
unit cell along the $b$ and $c$ axes leading to 16 different
magnetic sublattices in the magnetic unit cell, as shown in
Fig.~\ref{Fig-magn-structure}(a). Using this extended unit cell we
calculated the interaction matrix $\eta$ in the mean-field
approximation; the eigenvectors of $\eta$ are basis vectors of the
magnetic ordering and the eigenvalues are the corresponding energy
levels. Diagonalization of $\eta$ gives the two lowest energy
levels $\lambda_{\pm}=-J \pm J_{ac}$, assuming weakly-coupled AF
chains running along the $b$-axis ($J>0$ and $J \gg
J_{ab},J_{ac},J_{bc}$). For AF exchange between spins 1 and 2
($J_{ac}>0$) the ground state has energy $\lambda_{-}=-J-J_{ac}$
and there are 3 degenerate eigenvectors: (1) ordering of the
$y$($b$) spin components in the irreducible representation
$\Gamma^{10}$ (Eq.~\ref{rep10}), (2) ordering of the
$z$($c$)-components also in $\Gamma^{10}$ or (3) ordering of the
$x$($a$)-components in $\Gamma^{9}$ (Eq.~\ref{rep9}).\par

Upon including anisotropy effects this three-fold degeneracy is
lifted favoring ordering along the $b$-axis ($b$ is a common easy
axis for all spins as shown by paramagnetic susceptibility
measurements\cite{Duxbury,Pnam}). The determined magnetic
structure shown in Fig.~\ref{Fig-magn-structure} is indeed in the
$\Gamma^{10}$ representation with the largest spin moment along
the easy $b$-axis and a small moment along $c$, also in
$\Gamma^{10}$. The observed confinement of the ordered moments to
the ($b,c$) plane can also be understood on energetic grounds:
according to Eq.~\ref{rep10} ordering of spins along $a$ in
$\Gamma^{10}$ would have parallel spins on sites 1 and 2, which
would be energetically unfavored by the AF interchain couplings
$J_{ac}$ (the ordering of the $a$ components can not belong to
another representation, say $\Gamma^{9}$, because the full
Hamiltonian including all anisotropy terms respects the symmetry
of the crystal structure and therefore does not mix spin orderings
from different irreducible representations).\par

For a magnetic ordering of the $b$ and $c$ spin components in the
$\Gamma^{10}$ representation the spin configuration is not unique,
but instead four distinct domains are possible, all with the same
mean-field exchange energy. Those four domains are shown in
Fig.~\ref{Fig-magn-structure} and the degeneracy arises because
the $\Gamma^{10}$ representation of the $b$ and $c$ components is
two-dimensional (sites 1 and 3 and independent).\par

Interestingly, at the mean-field level the ground-state energy
does not depend on the couplings between sublattices (1,2) and
(3,4). This is a general result for any antiferromagnetic ordering
along $b$, i.e. $\bbox{k}$=(0,1/2,$l$) as can be easily seen by
inspecting Fig.~\ref{Fig-magn-structure}(a): each spin on the
second group of sublattices interacts with pairs of antiparallel
spins from the first group such that the net interactions cancel
out. For example, spin 4 interacts through $J_{bc}$ with the pair
of antiparallel spins 1 and 9 (5 and 13). Similarly, spin 4 at
$x$$\sim$ 0.25 interacts through $J_{ab}^{\prime}$ with the pair
of antiparallel spins 6 and 14 at $x$$\sim$0.75 ($J_{ab}$ with
antiparallel spins type 6 and 14 at $x$$\sim$$-0.25$).\par

The observed ordering at $\bbox{k}$=(0,1/2,1/2) thus consists of
two interpenetrating magnetic lattices that are non-interacting at
the mean-field level. One lattice containing atoms
$1,\,2,\,5,\,6,\,9,\,10,\,13,\,14$, and the other has atoms
$3,\,4,\,7,\,8,\,11,\,12,\,15,\,16$ in the magnetic unit cell.
Since those two global lattices are related by inversion symmetry
($\bar{1}$ at the center of the chemical unit cell) the
intra-sublattice interactions are identical and so the ordered
magnetic moment is expected to be the same for both, as assumed in
the analysis of the magnetic Bragg peaks in
Section~\ref{Sec_results} that gave good agreement with the
experiment.\par

The observed small alternating tilt $\phi=\pm$15(5)$^{\circ}$ of
the magnetic moments away from the $b$-axis indicates an effective
local spin anisotropy. We do not have an explanation for its
origin but propose that it may arise when the Heisenberg
interchain exchange between sites with rotated XY planes (such as
$J_{ac}$ between spins 1 and 2 in Fig.~\ref{crystal_structure}) is
projected onto the lower-lying Kramers doublet of effective
spin-1/2.\par

\subsection{Absolute magnitude of the ordered moment}
\label{Subsec_magnitude} The magnitude of the ordered moment at
$T=80\;\mathrm{mK}$ was determined by comparing the nuclear and
the AF Bragg peak intensities (see Appendix C). Using $A_{\rm
exp}$ and Eq.~\ref{Eq_int} the resulting ordered moment at
$T=80\;\mathrm{mK}$ is $m_0=1.7(4)\mu_B$ (zero field).\par

The lower-lying Kramers doublet of Co$^{2+}$ ions with effective
spin-1/2 has anisotropic $g$-values, $g_{x,y}=2g$ and $g_z=g$.
Here $g$=2.4 is the isotropic $g$-value of the underlying spin-3/2
and was determined from high-temperature paramagnetic
susceptibility measurements \cite{Smit,McElearney,Figgis}. This
implies that the available moment in the XY plane is
$g_{x,y}\mu_B/2$=2.4 $\mu_B$. From the present diffraction
experiments the estimated ordered moment along the in-plane axis
$b$ common to all spins is $m_0\cos\phi$=1.6(4) $\mu_B$, clearly
smaller than the available full moment in the XY plane (the
ordered moment along the $c$-axis is very small and is a mixture
between longitudinal and transverse parts). The reduction of the
ordered moment from the full available value indicates strong
zero-point fluctuations in the ground state.\par

\subsection{Magnetization curve}

\label{Subsec_field_dependence} In this section we consider a
minimal magnetic Hamiltonian consistent with the crystal field and
magnetization data on Cs$_2$CoCl$_4$ and relate the observed
magnetic phase diagram to the phenomenology of anisotropic magnets
in noncommuting fields. We analyze the magnetization curve in
terms of a spin-3/2 Hamiltonian appropriate for the Co$^{2+}$ ions
and identify the observed crossover at $H_m$ with near-saturation
of the lower-lying spin doublet. We then estimate the
lower-doublet magnetization curve
(Fig.~\ref{Fig_lower_doublet_magnetization}) and discuss it in
terms of an effective $S$=1/2 XXZ Hamiltonian in noncommuting
field.

\subsubsection{Full Hamiltonian}
A minimal Hamiltonian for the spin $\tilde{S}$=3/2 Co$^{2+}$ ions
that includes the 1D exchange and crystal field effects is
\begin{equation}
    \tilde{{\mathcal H}}=\sum_i I \tilde{\bbox{S}}_i \cdot \tilde{\bbox{S}}_{i+1}
    +D\left( \tilde{S}_i^{z} \right)^2 -g\mu_B H_z \tilde{S}_i^{z} -g\mu_B H_x \tilde{S}_i^x,
    \label{Eq_full_Ham}
\end{equation}
where $I$ is the nearest-neighbor (isotropic) exchange interaction
along the 1D chains and $D$ is the easy-plane anisotropy energy
(perpendicular to the local $z$-axis). The last two terms in
Eq.~\ref{Eq_full_Ham} are the Zeeman energy in longitudinal
($H_z=H\sin\beta$) and transverse ($H_z=H\sin\beta$) fields, where
$\beta$ is the angle between the field direction and the XY plane
(for $H$$\parallel$$a$ this is equal to the angle between the
local $z$-axis and the $c$-axis in Fig.~\ref{crystal_structure}).


In the absence of magnetic fields ($H=0$) and in the limit of
large local anisotropy ($D \gg I$) only the lower-lying doublet
$|\pm\frac{1}{2}\rangle$ contributes to the low-energy dynamics
and in this subspace the degrees of freedom can be
described\cite{Algra} by an effective spin $S$=1/2 XXZ Hamiltonian
(Eq.~\ref{Eq_Hamil_CsCoCl}) with exchange $J=4I$ and anisotropy
parameter $\Delta$=0.25.

\subsubsection{XXZ chain in a longitudinal (commuting) field}

The physics of the XXZ model (Eq.~\ref{Eq_Hamil_CsCoCl}) is well
understood in the limit $\Delta$=0 when it is equivalent to a 1D
free fermion gas. Longitudinal fields ($H \parallel z$) act as
chemical potential filling up the quasiparticle band. The
magnetization is directly related to the filling factor and at
finite temperature $T$ is given by \cite{Katsura}
\begin{equation}
M(H,T)=\frac{M_{1z}}{\pi}\int_0^{\pi} d\omega
\tanh\frac{g_z\mu_B(H-H_{1z}\cos\omega)}{2k_BT},
\label{Eq_magnetization}
\end{equation}
with the $T$=0 result
$M(H)=2M_{1z}\sin^{-1}\left(H/H_{1z}\right)/{\pi}$, $H \leq
H_{1z}$. Fig.~\ref{Fig_lower_doublet_magnetization} shows a
typical magnetization curve plot at $T$=0 (dashed curve) and
finite $T$ (solid line) where the approach to saturation is
rounded off by thermal fluctuations. The saturation magnetization
$M_{1z}=g_z\mu_B/2$ (per spin) is reached at the critical field
(for finite $\Delta$) $H_{1z}=J(1+\Delta)/g_z\mu_B$. The
zero-field, $T=0$ longitudinal susceptibility is reduced compared
to the semi-classical value due to quantum fluctuations and for
$\Delta$=0.25 is calculated as \cite{Muller84}
$\chi_z(0)$$\simeq$0.237$g_z^2\mu_B^2/J$. Using the proposed
values for the exchange interactions in Cs$_2$CoCl$_4$,
$J$=0.23(1) meV, $g_z$=2.4 the longitudinal critical field is
estimated as $H_{1z}$=2.1(1) T.

\subsubsection{XXZ chain in a transverse (noncommuting) field}

Transverse fields ($H||x$) have a very different effect compared
to longitudinal fields because they : (1) break the spin
rotational symmetry from U(1) to Ising, and (2) do not commute
with the exchange terms. In low fields this produces perpendicular
long-range antiferromagnetic order (spin-flop phase)
\cite{Kurmann81,Kurmann82}. The non-commutation of the field
introduces fluctuations that become strong enough above a critical
field to suppress the antiferromagnetic order and induce a
transition to a spin liquid state where the spin moment is not yet
saturated along the field and the finite perpendicular spin
components have exponentially-decaying correlations. A typical
ground state phase diagram is shown in
Fig.~\ref{Fig_theoretical_phase_diagram}. The critical disordering
field is estimated \cite{Kurmann82} at $H_{1x}^{\prime}\simeq
J(3+\Delta)/2g_x\mu_B$ whereas exact diagonalizations of finite
chains\cite{Kurmann81} suggest that near-saturation of the
magnetization occurs at much higher fields, similar to the
classical saturation value $H_{1x}=2J/g_x\mu_B$. For
Cs$_2$CoCl$_4$ those estimates give $H_{1x}^{\prime}$
$\simeq$1.3(1) T and $H_{1x}$=1.7(1) T. The zero-field, $T=0$
susceptibility is again reduced compared to the semiclassical
value due to quantum fluctuations and for the XY chain is
calculated as \cite{Muller84} $\chi_{x}(0)\simeq 0.075
g_x^2\mu_B^2/J$.

\subsubsection{Full Hamiltonian in a longitudinal field}

For the $\tilde{S}$=3/2 Hamiltonian in Eq.~\ref{Eq_full_Ham} with
well-separated energy scales for the lower
$|\pm\frac{1}{2}\rangle$ and higher $|\pm\frac{3}{2}\rangle$
doublets a cross-over in behavior is expected in applied magnetic
field from a low-field region where only the lower doublet
participates (or both doublets depending on the field direction)
to a higher field region where the lower doublet is saturated
and only the higher-doublet contributes to the magnetization.

Longitudinal fields ($H \parallel z$) do not mix higher-doublet
states until very large fields of the order the inter-doublet
separation energy $H_{2z}\simeq2D/g_z\mu_B$. At low fields only
the lower-doublet states contribute and the physics is that of the
XXZ model in longitudinal fields. Above the lower-doublet
saturation at $H_{1z}$ the magnetization shows a plateaux at
$M_{1z}=g_z\mu_B/2$ (per spin) stable up to fields around $H_{2z}$
when the magnetization increases again by mixing in states from
the higher-lying doublet $|\pm\frac{3}{2}\rangle$ to finally reach
the full spin value of $3/2g\mu_B$. The upper longitudinal
critical field for Cs$_2$CoCl$_4$ is estimated at $H_{2z}$=9(1) T.

\subsubsection{Full Hamiltonian in a transverse field}

Transverse fields ($H$ $\parallel$ $x$) in Eq.~\ref{Eq_full_Ham}
have finite matrix elements between the two doublets and thus mix
higher-doublet states into the ground state (of order
$g\mu_BH/2D$) at any finite field.\cite{Smit} Both doublets
participate at low fields (large susceptibility) and above a
cross-over field the lower doublet is near saturated and only the
higher doublet contributes (small susceptibility). Such a behavior
is evident in earlier magnetization measurements\cite{Smit,Pnam}
on Cs$_2$CoCl$_4$ in fields along the $b$-axis (entirely
transverse) where the magnetization increases rapidly at low
fields with a large susceptibility then crosses over above $\sim$3
T to a region where the magnetization increases much slower to
approach saturation (3.6 $\mu_B$) at fields $>$16 T.

\subsubsection{Comparison with the observed (total) magnetization}

The $a$-axis magnetization shown in Fig.~\ref{Fig_moments} is in
broad agreement with the expected behavior for mixed longitudinal
and transverse fields on the full $\tilde{S}$=3/2 Hamiltonian in
Eq.~\ref{Eq_full_Ham}: a rapid increase is observed at small
fields (both doublets contribute) followed by a cross-over to a
much slower increase at higher fields (only higher-doublet states
contribute). The cross-over field $H_m$=2.52$\pm$0.06 T ($T$=80
mK) is identified with near-saturation of the lower doublet. This
field appears to be larger than the estimated near- or
full-saturation fields for purely transverse $H_{1x}$ or purely
longitudinal fields $H_{1z}$, possibly due to either the
approximations used in estimating $H_{1x}$ solving
Eq.~\ref{Eq_full_Ham} assuming decoupled lower and higher
doublets, or other terms in the Hamiltonian, such as
interchain exchanges not included explicitly here.

Fig.~\ref{Fig_lower_doublet_magnetization} shows the partial
lower-doublet magnetization assuming the contribution to the
magnetization from the higher doublet states can be approximated
by a constant susceptibility (as shown in Fig.~\ref{Fig_moments}).
This assumption is consistent with calculations in the single-site
approximation following Ref.~\onlinecite{Smit}, which predict that
the higher-doublet susceptibility for fields tilted at
$\beta$=45$^{\circ}$ is relatively small (compared to
$\chi_{x,z}(0)$ of the lower doublet) and decreases smoothly by
only $\sim$25$\%$ between zero and 5 T. Putting the magnetization
values in Fig.~\ref{Fig_lower_doublet_magnetization} on an
absolute scale gives a lower-doublet saturation moment
$M_s$=1.7(4) $\mu_B$, consistent with typical values expected for
fields applied at an angle to the $z$-axis
$M_s=\sqrt{g_x^2\cos^2\beta+g_z^2\sin^2\beta}\mu_B/2\simeq$1.9
$\mu_B$ for $\beta$=45$^{\circ}$.

\subsubsection{(Partial) magnetization of the XXZ chain}

Since no detailed predictions are available for the magnetization
curve of XXZ chains in mixed longitudinal and transverse fields we
compare the results with the generic analytic form given in
Eq.~\ref{Eq_magnetization}. This is valid strictly only for XY
chains in longitudinal fields, but it provides a simple analytic
form to parametrize the data and extract an effective saturation
field $H_s$, the saturation magnetization and an effective
``temperature'' $\tilde{T}$, which is a measure of the
fluctuations causing the rounded approach to saturation.

\begin{figure}
\begin{center}
  \includegraphics[height=6cm,bbllx=77,bblly=265,bburx=485,
  bbury=568,angle=0,clip=]{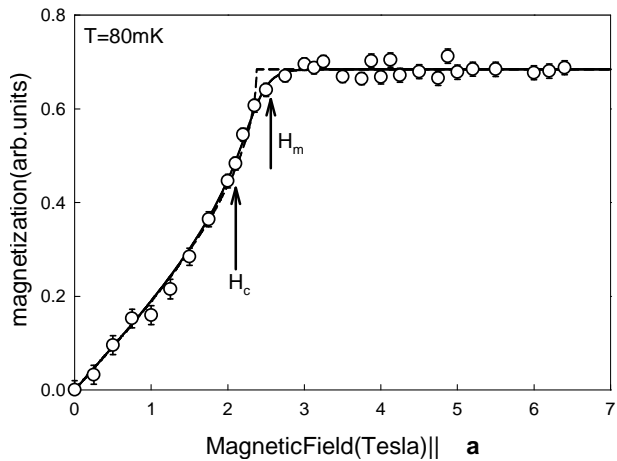}
  \caption{Partial magnetization from lower-doublet states alone as
  a function of applied field along $a$. The data is obtained
  from the observed total magnetization by subtracting an estimate of
  the higher-doublet contribution as indicated in
  Fig.~\protect\ref{Fig_moments} (dashed line) and described
  in the text. $H_c$=2.1 T is the critical field where the
  antiferromagnetic order in the ($b,c$) plane is suppressed
  (see Fig.\protect\ref{Fig_field_1AF_80mK}) and $H_m$ is the cross-over field
  above which the (lower-doublet) magnetization is nearly (within 5\%)
  saturated. The solid line is a fit to the generic form in
  Eq.~\protect\ref{Eq_magnetization} (the dashed line shows
  this calculation at zero temperature) as described in the text.}
  \label{Fig_lower_doublet_magnetization}
\end{center}
\end{figure}

The extracted saturation field $H_s$=2.37(3) T (80 mK$<$$T$$<$ 250
mK) is significantly larger than the critical field $H_c$=2.10(4)
T where the antiferromagnetic order disappears, giving further
support for the existence of an intermediate phase between the
antiferromagnetic order and the nearly fully-polarized phase as
indicated in the phase diagram in Fig.~\ref{Fig_phase-diagram}.
The fitted ``temperatures'' $\tilde{T}$ are systematically larger
than the true measuring temperatures $T$ ($\tilde{T}$=200(50) mK
for the $T$=80 mK data, fit shown by solid line in
Fig.~\ref{Fig_lower_doublet_magnetization}) indicating more
fluctuations in the system than can be accounted for by
temperature alone. A source of those fluctuations can be the
noncommuting fields that create disorder effects at large fields.
At small noncommuting fields the dominant effect is breaking
the spin rotational symmetry which has the consequence
of promoting long-range order in a spin-flop phase. This is also
in agreement with the experiments, which observed that low fields
stabilize the perpendicular antiferromagnetic order.\par

\subsection{Further studies}

\label{Subsec_further_studies} The measurements presented here have
highlighted a potentially very significant field induced phase transition
in Cs$_2$CoCl$_4$. The field dependence of the order
correlates well with the expectations of a quantum magnet
driven through a quantum critical point by a noncommuting field.
Essential to this picture is the identification of the disordered phase with
a gapped spin liquid state. Further studies using NMR,
magnetic susceptibility, and heat capacity are called for
to investigate whether a spin gap really does open
above the critical field of 2.10 T. It would also
be very interesting to look for any evidence of spin glass
behavior induced by the field. Inelastic neutron scattering measurements
of the excitations in a field should be particularly
revealing and we plan to make such measurements in the near future.

\section{Conclusions}
\label{Sec_conclusions} In conclusion, single-crystal neutron
diffraction was used to determine the magnetic ordering as a
function of noncommuting applied field in the quasi-1D spin-1/2
XY-like antiferromagnet Cs$_2$CoCl$_4$. In zero field long-range
order with wavevector $\bbox{k}$=(0,1/2,1/2) was found below
$T_N$=217(5) mK. The magnetic structure was determined using group
theory and has spins ordered antiferromagnetically along the
chains with moments confined to the ($b,c$) plane. A domain
structure was found with adjacent chains in different phases.
Possible mechanisms promoting this commensurate order were
discussed. The observed reduction in the ordered moment was
attributed to zero-point fluctuations in the ground state.\par

Magnetic fields applied along the $a$-axis were found to initially
stabilize the perpendicular AF order and form a spin-flop phase.
This structure becomes unstable at high fields where a transition
occurs to a phase with no LRO in the ($b,c$) plane. Measurements
of the ferromagnetic component found that near-saturation of the
moments occurs only at much higher fields. The phase in-between
the spin-flop and saturated phases $2.10<H_{SL}<2.52$ T
$\parallel$ $a$ has been proposed to be a spin liquid state
disordered by the strong quantum fluctuations arising from the
applied field noncommutability with the exchange Hamiltonian.\par

\begin{acknowledgments}
We would like to thank R.~A. Cowley, F.~H.~L. Essler and M.
Meissner for stimulating discussions. Preliminary work for the
characterization of single crystals was performed at Chalk River
Laboratories in Canada. Financial support for the experiments was
provided by the EPSRC and by the EU through the Human Potential
Programme under IHP-ARI contract HPRI-CT-1999-00020. ORNL is
managed for the U.S. D.O.E. by UT-Battelle, LLC, under Contract
No. DE-AC05-00OR22725. One of the authors (M.K.) was supported by
the Swiss National Science Foundation under Contract No.
83EU-053223.
\end{acknowledgments}

\section*{Appendix}
\subsection{Group theory analysis}
\label{Subapp_group_theory} Using the space group symmetry of the
crystal structure we identify allowed basis vectors for a magnetic
structure with the observed wavevector $\bbox{k}$=(0,1/2,1/2).
This is done by determining the irreducible representations and
eigenvectors of the little group $G_{\bbox{k}}$ of symmetry
operations that leave the wavevector $\bbox{k}$ invariant.

We start by considering the symmetry elements of the \textit{Pnma}
space group of ${\rm Cs_2CoCl_4}$:
\begin{equation}
    \{ 1,\;\overline{1},\;2_x,\;2_y,\;2_z,\;m_{xy},\;m_{xz},
    \;m_{yz} \}\, ,
\end{equation}where $1$ is the identity operator, $\overline{1}$
is the inversion at the origin, $2_{\alpha}$ denotes a
180$^{\circ}$ screw axis along direction $\alpha=x, y $ or $z$
(180$^{\circ}$ rotation followed by a translation with half a unit
cell along axis $\alpha$) and where throughout this group-theory
section the axes $x$, $y$ and $z$ refer to the crystallographic
directions $a$, $b$ and $c$. $m_{\alpha \beta}$ is a glide plane
containing axes $\alpha$ and $\beta$. The group is nonsymmorphic
because the group elements $\{R|\bbox{a}\}$ consist of an
operation $R$ followed by a translation $\bbox{a}$ equal to half a
direct lattice vector. The experimentally observed ordering
wave-vector $\bbox{k}=(0,1/2,1/2)$ is invariant under all these
operations such that the little group $G_{\bbox k}$ contains all
of the above elements.\par

The representations of the little group $G_{\bbox k}$ are given by
$\exp(-i {\bbox k}{\bbox a})\; \Gamma^{\alpha}(R_{\bbox k})$. The
ordering in ${\rm Cs_2CoCl_4}$ is a special case because the
crystal symmetry is nonsymmorphic and the ordering wave-vector
lies on the Brillouin zone boundary. The representations of
elements $R_{\bbox k}$ of the space group with subsequent
translation $\bbox{t}=(0,n,m)$ such that $n+m={\rm odd}$ have a
different sign to those of operations $R_{\bbox k}$ alone, and
this leads to additional irreducible representations. They can be
found by adding new group elements to the little group $G_{\bbox
k}$,\cite{Heine} which are the original elements with an
additional translation $\bbox{t}$. We call the additional
translations
\begin{equation}
    \{ 1^{t},\;\overline{1}^{t},\;2_x^{t},\;2_y^{t},\;2_z^{t},
    \;m_{xy}^{t},\;m_{xz}^{t}, \;m_{yz}^{t} \}\, .
\end{equation}Thus the little group $G_{\bbox k}$ consists of a
total of $16$ elements. We determined the classes and the
character table of this group and these are shown in
Table~\ref{basis_vector_table}. The group consists of $10$
different classes and therefore has $10$ irreducible
representations. Only two of the representations fulfill the
necessary condition that $\chi(1) = -\chi(\overline{1}^{t})$ (this
condition follows directly from the prefactor of the
representation which changes sign under a translation by
$\bbox{t}$). The experimentally observed ordered magnetic
structure is thus associated with one of these two
representations.\par

\begin{table*}\begin{ruledtabular}
\begin{tabular}{ccccccccccccc}

 & &$1$ & $1^t$ & $2_x$ & $2_x^t$ & $\overline{1}/\overline{1}^t$
  & $2_y/2_y^t$ & $2_z/2_z^t$ & $m_{xy}/m_{xy}^t$ & $m_{xz}/m_{xz}^t$
  & $m_{yz}/m_{yz}^t$ &\\ \hline
& $\Gamma^1$ & 1 & 1 &  1 &  1 &  1 &  1 &  1 &  1 &  1 &  1 &\\ &
$\Gamma^2$ & 1 & 1 &  1 &  1 & -1 & -1 &  1 &  1 & -1 & -1 &\\ &
$\Gamma^3$ & 1 & 1 &  1 &  1 & -1 &  1 & -1 & -1 & -1 &  1 &\\ &
$\Gamma^4$ & 1 & 1 &  1 &  1 &  1 & -1 & -1 & -1 &  1 & -1 &\\ &
$\Gamma^5$ & 1 & 1 & -1 & -1 &  1 &  1 & -1 &  1 & -1 & -1 &\\ &
$\Gamma^6$ & 1 & 1 & -1 & -1 &  1 & -1 &  1 & -1 & -1 &  1 &\\ &
$\Gamma^7$ & 1 & 1 & -1 & -1 & -1 &  1 &  1 & -1 &  1 & -1 &\\ &
$\Gamma^8$ & 1 & 1 & -1 & -1 & -1 & -1 & -1 &  1 &  1 &  1 &\\ &
$\Gamma^9$ & 2 &-2 & -2 &  2 &  0 &  0 &  0 &  0 &  0 &  0 &\\ &
$\Gamma^{10}$& 2 &-2 &  2 & -2 &  0 &  0 &  0 &  0 &  0 &  0 &\\
\end{tabular}
\caption{Irreducible representation of the group $G_{\bbox k}$.}
\label{basis_vector_table}
\end{ruledtabular}
\end{table*}
The eigenvectors $\phi^{\lambda}$ of the irreducible
representations $\Gamma^{\lambda}$ were determined using the
projector method.\cite{Heine} They are given by
\begin{equation}
    \phi^{\lambda}=\sum_g \chi^{\lambda}(g) g(\phi),\, ~~~~~~\lambda=1 \ldots 10,
\end{equation}where $g$ is an element of the little group and
$\phi$ is any vector of the order parameter space. For
$\bbox{k}$=(0,0.5,0.5) the magnetic unit cell is doubled along the
$b$ and $c$ axes, and the order parameter space is a
$48$-dimensional axial vector because the magnetic unit cell
contains $16$ magnetic moments as indicated in
Fig.~\ref{Fig-magn-structure})(a) and each has three space
components. We obtain
\begin{eqnarray}\label{rep9}
    &\phi^{9}=(m_{1x},m_{1y},m_{1z},-m_{1x},m_{1y},m_{1z},
    &\nonumber\\ &m_{3x},m_{3y},m_{3z},-m_{3x},m_{3y},m_{3z},
    &\nonumber\\&-\bbox{m}_1,-\bbox{m}_2,-\bbox{m}_3,-\bbox{m}_4,
    &\nonumber\\&-\bbox{m}_1,-\bbox{m}_2,-\bbox{m}_3,-\bbox{m}_4,
    &\nonumber\\&\bbox{m}_1,\bbox{m}_2,\bbox{m}_3,\bbox{m}_4)&
\end{eqnarray} for representation $\Gamma^{9}$. $m_{i \alpha}$ is
the component $\alpha$ of the magnetic moment $i$ with $1\leq i
\leq 16$. For better readability, in the above equation the
individual components of the magnetic moments 5 to 16 were omitted
and the magnetic moments were written as vectors, i.e.
$\bbox{m}_5=-\bbox{m}_1=(-m_{1x},-m_{1y},-m_{1z})$. For
representation $\Gamma^{10}$ we obtain
\begin{eqnarray}\label{rep10}
    &\phi^{10}=(m_{1x},m_{1y},m_{1z},m_{1x},-m_{1y},-m_{1z},
    &\nonumber\\ &m_{3x},m_{3y},m_{3z},m_{3x},-m_{3y},-m_{3z},
    &\nonumber\\&-\bbox{m}_1,-\bbox{m}_2,-\bbox{m}_3,-\bbox{m}_4,
    &\nonumber\\&-\bbox{m}_1,-\bbox{m}_2,-\bbox{m}_3,-\bbox{m}_4,
    &\nonumber\\&\bbox{m}_1,\bbox{m}_2,\bbox{m}_3,\bbox{m}_4)\, .&
\end{eqnarray}The ordered magnetic structure is thus doubly degenerate
in each of the three spin components given by the dimension 2 of
these two representations.\par

A magnetic structure in the $\Gamma^{10}$ irreducible
representation can occur in several different domains. Assume
spins are confined to the ($b,c$) plane in a typical configuration
shown in Fig.\ref{Fig-magn-structure}(a) called domain A1. One can
construct domain A2 in Fig.\ref{Fig-magn-structure}(c) by
reversing the $b$($y$)-components of all spins and keeping the
$c$($z$)-components unchanged. Domain B1 in
Fig.\ref{Fig-magn-structure}(b) is obtained from domain A1 by
reversing the spins on sites 3, 4 (and 7, 8, 11, 12, 15, 16) and
using the same rule one transforms domain A2 in
Fig.\ref{Fig-magn-structure}(c) into B2 in
Fig.\ref{Fig-magn-structure}(d). Domains A1 and A2 are
indistinguishable from each other in a neutron scattering
 experiment because they have the same magnetic
structure factor and domains B1 and B2 are also indistinguishable.
However, an A-type domain (either A1 or A2) has a different
structure factor from a B-type (either B1 or B2) domain.\par
\subsection{Zero-field magnetic structure}\label{Subapp_mag_structure}

The integrated intensity of magnetic Bragg peaks is related to the
structure factor of the magnetic ordering through \cite{Squires}
\begin{equation}\label{Eq_int}
    I(\bbox{Q})=\left(\frac{\gamma r_0}{2\mu_B}\right)^2 N_m \frac{(2\pi)^3}{V_{m0}}
    \Phi \frac{|f(\bbox{Q})|^2}{\sin(2\Theta)} |{\bbox
    F}_{\bot}(\bbox{Q})|^2\, ,
\end{equation} where $f(\bbox{Q})$ is the magnetic form factor for Co$^{2+}$ ions
\cite{IntTables}. $\Phi$ is the flux of incident neutrons, $N_m$
and $V_{m0}$ are the number and the volume of the magnetic unit
cells, $\gamma=1.913$ and $r_0=2.818\cdot10^{-15}\;\mathrm{m}$.
$I(\bbox{Q})$ is the total integrated intensity of a Bragg
reflection measured in the ($\Psi$,$2\Theta$) plane and
$1/\sin(2\Theta)$ is the Lorentz correction factor that arises
because intensity is measured as a function of angular
coordinates. ${\bbox F}_{\bot}(\bbox{Q})$ is the component of the
magnetic structure factor perpendicular to the scattering wave
vector and is defined as
\begin{equation}\label{Eq_fperp}
    {\bbox F}_{\bot}(\bbox{Q}) = {\bbox F}(\bbox{Q}) -
    ({\bbox F}(\bbox{Q}) \cdot \bbox{\hat{Q}})\;\bbox{\hat{Q}},
\end{equation} where $\bbox{\hat{Q}}$ is the normalized wave-vector
transfer.\par The magnetic structure factor is defined as
\vspace{0.5cm}
\begin{equation}\label{Eq_f}
    {\bbox F}(\bbox{Q}) = \sum_{i=1}^{16} \bbox{m}_i \exp (-{\rm i}\;
    \bbox{Q} \cdot \bbox{d}_i)
\end{equation}
where $\bbox{Q}$ is the wave-vector transfer in the experiment,
$\bbox{d}_i$ are the positions of the Co$^{2+}$-ions and the sum
is over all $16$ magnetic ions in the magnetic unit cell. For a
two-domain structure (A + B type) the magnetic Bragg peak
intensity can be written as
\begin{equation}\label{Eq_Ifit}
    {\bbox I}(\bbox{Q})=\alpha{\bbox I}_A(\bbox{Q}) +
    (1-\alpha){\bbox I}_B(\bbox{Q})\, ,
\end{equation}where $\alpha$ is the population of the A-domain
and $(1-\alpha)$ is the population of the B-domain.\par

\subsection{Absolute magnitude of the ordered moment}
\label{Subapp_magnitude} The magnitude of the ordered moment at
$T=80\;\mathrm{mK}$ was determined by comparing the nuclear and
the AF Bragg peak intensities. The intensity of a nuclear Bragg
peak is given as
\begin{equation}
    I(\bbox{Q})=N \frac{(2\pi)^3}{V_{0}} \Phi
    \frac{|F_N(\bbox{Q})|^2}{\sin(2\Theta)}\,
    \label{Eq_nuclint},
\end{equation}where $N$ and $V_{0}$ are the number and the
volume of the unit cells. $F_N(\bbox{Q})$ is the nuclear structure
factor and given as\cite{Squires}
\begin{equation}
    F_N(\bbox{Q})=\sum_{i}b_i \exp(-{\rm i}\, \bbox{Q} \cdot \bbox{d}_i)\,
    ,
\end{equation}where the sum runs over all elements in the nuclear
unit cell and $b_i$ is the elastic scattering
length\cite{IntTables} of atom $i$ in the unit cell. The nuclear
peaks used for calibration were (02$\bar{2}$), (033), (022) and
(03$\bar{3}$) and their observed relative integrated intensities
were consistent with the calculated structure factors to within
15$\%$. Multiple scattering and extinction corrections were
assumed to be negligible. The measured nuclear intensities gave
the overall scale factor for the intensities $A_{\rm exp}=N
\frac{(2\pi)^3}{V_0} \Phi$ in Eq~\ref{Eq_nuclint}. The magnitude
of the ordered magnetic moment was determined using $A_{\rm exp}$
and Eq.~\ref{Eq_int}. The resulting ordered moment at
$T=80\;\mathrm{mK}$ in zero field is $m_0=1.7(4)\mu_B$.\par

\bibliographystyle{prsty}

\end{document}